# MULTI-COLOR POLARIMETRY OF THE TWILIGHT SKY.
# THE ROLE OF MULTIPLE SCATTERING AS THE FUNCTION OF WAVELENGTH.

## O.S. Ougolnikov[1,2], and I.A. Maslov[2]

[1]*Astro-Space Center of Lebedev's Physical Institute, 117997, Profsoyuznaya 84/32, Moscow, Russia*
[2]*Space Research Institute, 117997, Profsoyuznaya 84/32, Moscow, Russia*

**Abstract.** This work is devoted to the study of solar light that experiences single and multiple scattering in the atmosphere at the twilight time. The work is based on polarimetric observations of the twilight sky that were conducted in 1997 and 2000 for different color bands. Basing on the results of observations, the contribution of single-scattered light depending on the wavelength for the light stage of twilight is obtained. The correlation of color and polarization properties of the twilight sky is investigated with an account of multiple scattering.

## 1. INTRODUCTION

Taking account of solar light that experiences multiple scattering in the Earth's atmosphere is one of the basic problems in the study of the atmosphere by photometric observations of the twilight sky. As early as 1923, V.G. Fesenkov called attention to the essential contribution of multiple scattering in his first work devoted to the analysis of twilight phenomena [1]. However, as it was correctly noted by G.V. Rozenberg [2], the extreme complexity of constructing a theory of multiple scattering forced many authors to disregard this effect or introduce various, sometimes fairly crude assumptions concerning its properties. A comparative analysis of a great number of works made by Rozenberg [2] revealed an extremely wide range of results, from the possibility of completely neglecting multiple scattering to its essential predominance over single-scattered light. Sufficient errors of defined ratio of single and multiple scattering often lead to wrong explanations of well-known observed color and polarization properties of the twilight sky. Obviously, the ratio of single- to multiple-scattered light directly determines the efficiency and work range of the twilight method of atmospheric study, in which the multiple scattering is a peculiar kind of "noise" that must be subtracted from the total sky background brightness. This twilight method is effective for investigations of lower atmosphere vertical profile, especially atmospheric ozone and aerosol properties.

The main object of investigation of this work is the behavior of the ratio of single and multiple scattered light during the twilight period and its relation with observed photometric and polarimetric properties of the twilight sky. The work is based on the improved method of calculation of this ratio from the observational polarimetric data.

## 2. PROPERTIES OF SINGLE AND MULTIPLE SCATTERING

Let us consider the scheme of formation of single- and multiple-scattered light at the twilight time and its fundamental properties. We shall be only interested in points in the solar vertical. The brightnesses of these two fractions of twilight sky background are denoted by $J$ and $j$, respectively. The scheme of appearance of both components at the sunrise and sunset times is shown in Figure 1. The brightness of the single-scattered component for a fixed wavelength is given by the integral

$$J = const \int_0^\infty e^{-\tau_1(H_L,z,h)} n(H_{SC}(H_L,z,h)) \times$$
$$\times D(H_{SC},\gamma) sec(z-h) e^{-\tau_2(H_L,z,h)} dH_L \quad (1),$$

where integration is performed over $H_L$, the minimum altitude of the solar ray above the Earth's surface; $H_{SC}$ is the altitude of the point of light scattering; $z$ is the zenith distance for an observed point of the sky (which is positive in the glow region and negative in the opposite region of the sky); $h$ is the depth of the sun under horizon; $n$ is the number density of particles at the altitude $H_{SC}$; $D$ is the scattering function (indicatrix); $\gamma$ is the scattering angle; $\tau_1$ and $\tau_2$ are the optical depths along the ray paths before and after the scattering event, respectively. The quantity $\tau_2$, with a small correction, is equal to the vertical optical depth of the atmosphere multiplied by $sec\ z$.

The "twilight ray" model is very convenient in some cases. The essence of this model is that solar rays passing near the Earth's surface experience strong absorption, while rays passing at great heights are scattered very weakly in the upper, rarefied atmospheric layers. As a result, the main part of the single scattered component is formed in the so-called "twilight layer", which has a comparatively small thickness that does not sufficiently depend on the depth of the sun under the horizon and on the zenith distance of the observed point (except the case of large $\lambda$ and $h<0$, when the "twilight layer" grazes lowest layers of the atmosphere and even the surface of the Earth). The minimum altitude of the twilight layer above the ground grows comparatively slow with the sun's depth. Taking this fact into account, the integral in formula (1) can be replaced by the integrand at a certain point $H_{L0}(h)$ multiplied by the effective thickness of the layer, which can be included in the constant. In this case, the effective scattering altitude $H_{SC0}(z,h)$ will correspond to every observed point.

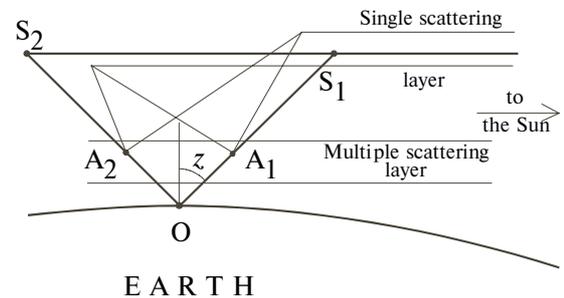

**Figure 1.** Single- and multiple-scattered light formation at the sunrise or sunset.

This method has already been used repeatedly, for example, by N.B. Divari [3], but in contrast to his work, we use the value of $H_{L0}$ corresponding to the average contribution to the integral instead of the $H_{L0}$ value corresponding to the maximum of the integrand. Owing to asymmetry of the twilight layer, these values differ by several kilometers.

Since the effective ray paths go at high altitudes (above 10 km), we can disregard refraction. Let us consider in greater detail the case $h=0$ (sun at the horizon). In this case (if the points close to the horizon are disregarded) $H_{SC0}=H_{L0}$, and equation (1) takes the simplified form

$$J = const \cdot e^{-\tau_1(H_{L0},z)} n(H_{L0}) D(H_{L0}, \frac{\pi}{2} - z) \times$$
$$\times sec\, z \cdot e^{-\tau_2(H_{L0},z)} = D(H_{L0}, \frac{\pi}{2} - z) f(z) \quad (2).$$

It should be noted that for $h>0$, the altitude of the twilight layer increases, $H_{SC0}(z)$ became $z$-dependent, and, as the sun descends under the horizon, this dependence becomes progressively steeper: the scattering in the glow region occurs in a lower place than in the opposite part of the sky.

We denote by $J_\perp(J_\parallel)$ the brightness of the single-scattered component in the polarization plane perpendicular (parallel) to the scattering plane. It is evident that equations (1) and (2) also holds for these quantities, with substitutions of the polarization indicatrices $D_\perp$ and $D_\parallel$, respectively. The quantities $j_\perp$ and $j_\parallel$ are introduced in a similar manner.

Instead of polarization coefficient we shall use in this work a more suitable quantity: the polarization ratio $K$, which is equal to the ratio of the sky brightnesses for polarization directions parallel and perpendicular to the scattering plane. The degree of polarization $p$ is connected with $K$ by simple relation $p=(1-K)/(1+K)$. For nonpolarized light $K$ is obviously equal to unity.

The scattering indicatrices $D_{\perp,\parallel}$ are the sum of air (Rayleigh) and aerosol indicatrices. From the Kabann-Rayleigh scattering matrix [2], one can readily obtain (accurate to constant) the air scattering indicatrices for the case of unpolarized outer emission:

$$D_\perp(\gamma) = 1 + \alpha; \quad D_\parallel(\gamma) = cos^2\, \gamma + \alpha \quad (3),$$

where $\alpha$ is 1/2 of the depolarization parameter in the Kabann-Rayleigh matrix. In accordance with [2], we assume $\alpha=0.03$. As for aerosol indicatrices, it is only known that, in contrast to the air indicatrices, they are asymmetric with respect to $\gamma=90°$, exhibiting an excess at small scattering angles.

Multiple scattering (as well as the scattering of light reflected from the Earth's surface), like single scattering, also take place in a certain layer, located, however, at much low altitudes, in the near-Earth atmospheric layers. Fesenkov [4] even called this component the "tropospheric" one. Its mean altitude varies at an extremely slow rate with the depth of the sun under horizon. Variation with zenith distance of the observed point is also insignificant. These facts form the basis for the assumption (which has been used repeatedly [4,5]) that the logarithmic derivatives at the symmetric points of the solar vertical for each $h$ value are equal to each other:

$$\frac{d\, ln\, j(z,h)}{dh} = \frac{d\, ln\, j(-z,h)}{dh} \quad (4).$$

However, as mean altitude of multiple scattering does vary, the equality cannot be integrated over a wide range of the sun's depths under the horizon (which is actually done in [5]), as will be shown below.

The effects related to the multiple scattering altitudes will disappear if we write equation (4) for two polarization directions and subtract one equation from another. We can then integrate the resulting equation, which will lead us to an expression relating the polarization ratios $q$ of multiple scattering at the symmetric points of the solar vertical:

$$q(z,h) = \theta(z) q(-z,h) \quad (5).$$

As we shall see later, the parameter $\theta$ for any $z$ is close to unity, and equation (5) describes the property of symmetry of multiple scattering polarization, which is quite natural if we take into account the similar polarization properties of the scattering of light at adjacent angles.

## 3. OBSERVATIONS AND PROPERTIES OF THE TWILIGHT SKY

First session of polarimetric observations of the twilight sky background was held in July — August 1997 at the Astronomical Observatory of Odessa University (Ukraine). The sky brightness and polarization were measured in a wide range of the sun's depth under the horizon: from −2° (sun above the horizon) to 20°. The measurements were made with an autoscanning twilight photometer [6] in the solar vertical in the range of zenith distances from −70° to +70°. The wavelength of measurements (356 nm) does not fall in the region of sufficient selective absorption of atmosphere gases, including ozone. The principal light absorption and scattering mechanism at this wavelength is Rayleigh scattering, to which scattering by particles of atmospheric aerosol may be added.

Second session of observations was conducted in July — August 2000 at South Laboratory of Moscow Sternberg Astronomical Institute (Crimea, Ukraine). Twilight sky background was registered by ST-6 CCD-camera with short-focus lens and polaroid. Maximum zenith distance was reaching about 15°. Exact axis direction and field of view were determined by the stars position at the nighttime images. At different dates observations were carried out in U, B, V and R color bands (with effective wavelengths about 360, 440, 550 and 700 nm, respectively), and at several evenings of August 2000 simultaneous observations in V and R bands were made. Effective wavelength of U band was near to first session observations wavelength. V and R bands, especially first one, fall into the region of Chappuis lines of ozone absorption. These lines are quite weak, but may have sufficient influence on $e^{-\tau_1}$ term in formulae (1) and (2) [7].

Figure 2 shows the dependencies of zenith polarization ratio $K$ by solar depth under horizon $h$ for some observation dates and colors. We can see that behavior of $K$ is principally the same for different wavelengths. At the sunset (or sunrise) polarization ratio is changing quite slowly and at $h>4$-$5°$ fast increasing (polarization weakening) starts. Here we should notice that at the same depths of the sun the behavior of twilight sky color is changing too, as we can see in Figure 3, slow reddening of the sky is replacing by its fast bluening [7]. These two well-known phenomena were repeatedly and hardly tried to explain as only single-scattered light effects [2], and in [8] the sky bluening was related with Chappuis ozone absorption, although we can observe this phenomenon even at the wavelengths outside the Chappius lines. As we shall see below, color and polarization changes have the common nature and can be easily explained if we take the multiple scattering effects into account.

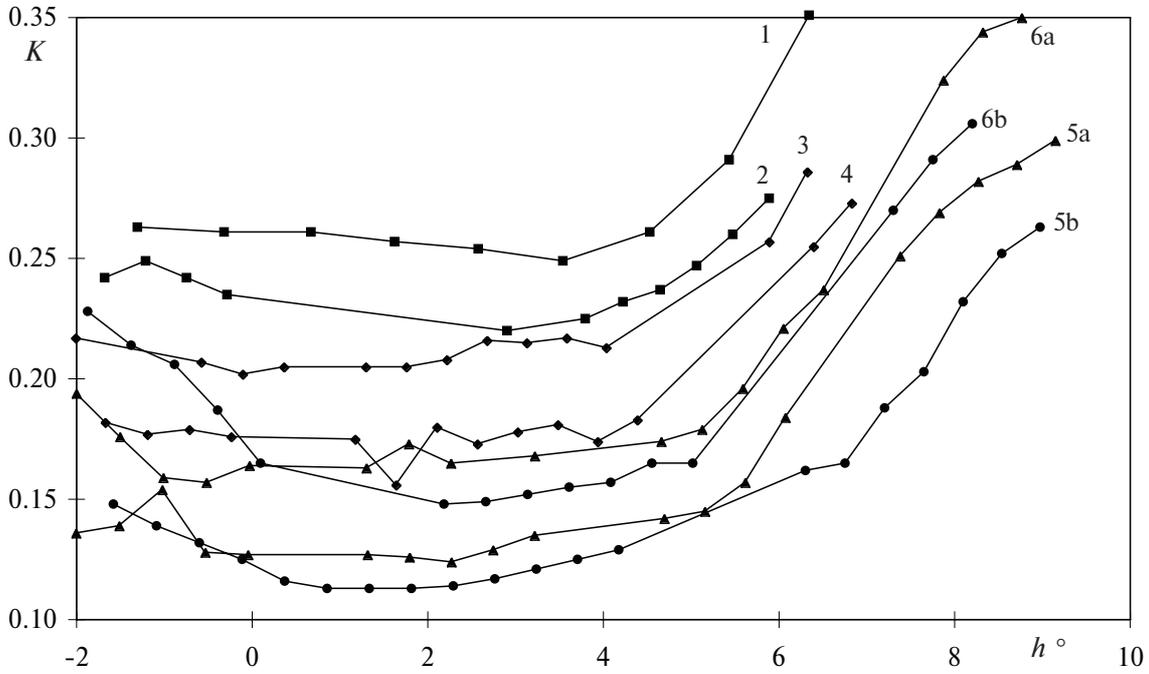

**Figure 2.** The polarization ratio $K$ of the twilight sky at the zenith depending on $h$ for some 1997 and 2000 observations (1 — evening, 31.VII.1997, 356 nm, 2 — evening, 17.VII.2000, U, 3 — morning, 26.VII.2000, B, 4 — morning, 28.VII.2000, B, 5a — evening, 4.VIII.2000, V, 5b — evening, 4.VIII.2000, R, 6a — evening, 6.VIII.2000, V, 6b — evening, 6.VIII.2000, R).

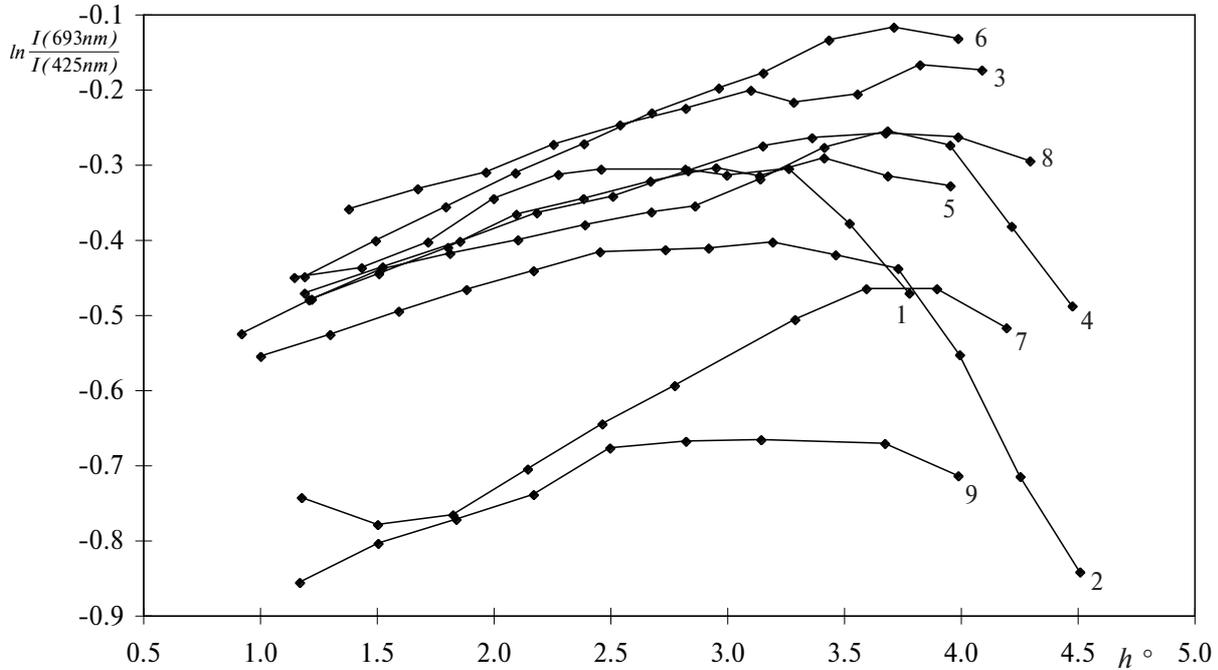

**Figure 3.** Color index of the twilight sky near the zenith ($z=-7.8°$) by 1994 observations at the Kuchino Astrophysical observatory [7] (1 — evening, 30.VI, 2 — morning, 8.VII, 3 — evening, 9.VII, 4 — morning, 10.VII, 5 — evening, 10.VII, 6 — evening, 13.VII, 7 — evening, 24.VII, 8 — evening, 28.VII, 9 — evening, 29.VII).

With the increasing of wavelength from U to V band the polarization ratio is decreasing (polarization getting stronger) in spite of the fact that depolarizing influence of aerosol scattering should rise at larger wavelengths. Polarization ratio in R band is near to one in V for the same observation day, as we can see in the Figure 2. As we may expect from this and shall see below, only in R band the influence of aerosol scattering is sufficient (we should note that observations were carried out only at good weather conditions), and at short wavelengths single aerosol scattering (at least for twilight ray altitudes corresponding to $h=0$) is small.

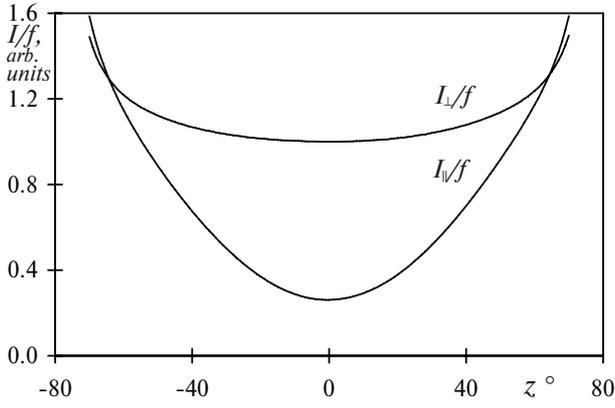

**Figure 4.** Symmetric profile of *(I/f)* quantity at the sunset of the evening, 31.VII.1997, 356 nm.

Last fact can be confirmed by the Figure 4, where the dependence of the value

$$\frac{I_{\perp(\parallel)}}{f(z)} = D_{\perp(\parallel)}(\frac{\pi}{2} - z) + \frac{j_{\perp(\parallel)}}{f(z)} \quad (6)$$

is shown for sunset moment of July, 31th, 1997, 356 nm. Function *f(z)* was calculated for this wavelength based on vertical optical depth measured by Bouguer's method for this date. We can see that this dependence is virtually symmetric with respect to the point $z=0$. Such situation is the same to other 1997 short-wavelength observation dates. Since the multiple scattered component cannot have an essential excess in the anti-solar vertical, we can conclude that the function *D* near the altitude $H_{L0}$ for $\lambda=356$ nm and $h=0$ turns out to be symmetric and, therefore, influence of aerosol scattering at this case is small.

Figure 5 shows the polarization ratio of the twilight sky *K* for different zenith distances as a function of *h*. Even these dependencies enable us, in principle, to assume at which depths of the sun under horizon the single scattering is completely lost against the background of the multiple scattering. This occurs at $h\approx10°$, because at greater depths the polarization ratios the symmetric points of the solar vertical become identical; i.e. property (5) with $\theta=1$, expected for multiple scattering, holds. The phenomenon of inverse polarization $(K>1)$ of the sky away from zenith is noteworthy. The reasons for its appearance will be discussed below.

At $h<10°$ sufficient asymmetry of polarization in the solar vertical appears, and sky background in the glow region is polarized stronger (*K* is less) than in the opposite one. As we will see below, it is related with appearance of strongly polarized single scattering component with an excess in the glow region due to variance of effective scattering heights.

The asymmetry of the single scattering relatively multiple one disappears near $h=0$ (in the case of weak aerosol scattering). At these values of *h* we can see than *K* is increasing with *h* for the points with $z<0$ and decreasing for $z>0$. This effect is related with a change of single scattering polarization due to change in angle of scattering, because polarization of the Rayleigh scattering reaches its maximum at $\gamma=90°$, which is corresponding to $z=0$ for $h=0$. The same phenomenon is clearly visible in the Figure 6, where the dependence of zenith distance of maximum polarization point on the solar depth under horizon is shown for simultaneous observation at V and R color bands in the evening of August, 6th, 2000. We can see that at small values of h this point is following the setting sun, but the velocity of such motion is less than in the case of pure Rayleigh single scattering (dashed line in the Figure 6). And at $h>5°$ (simultaneously with the polarization decrease and bluening of the sky) this point turns around to the zenith that can be explained only by domination of multiple scattering.

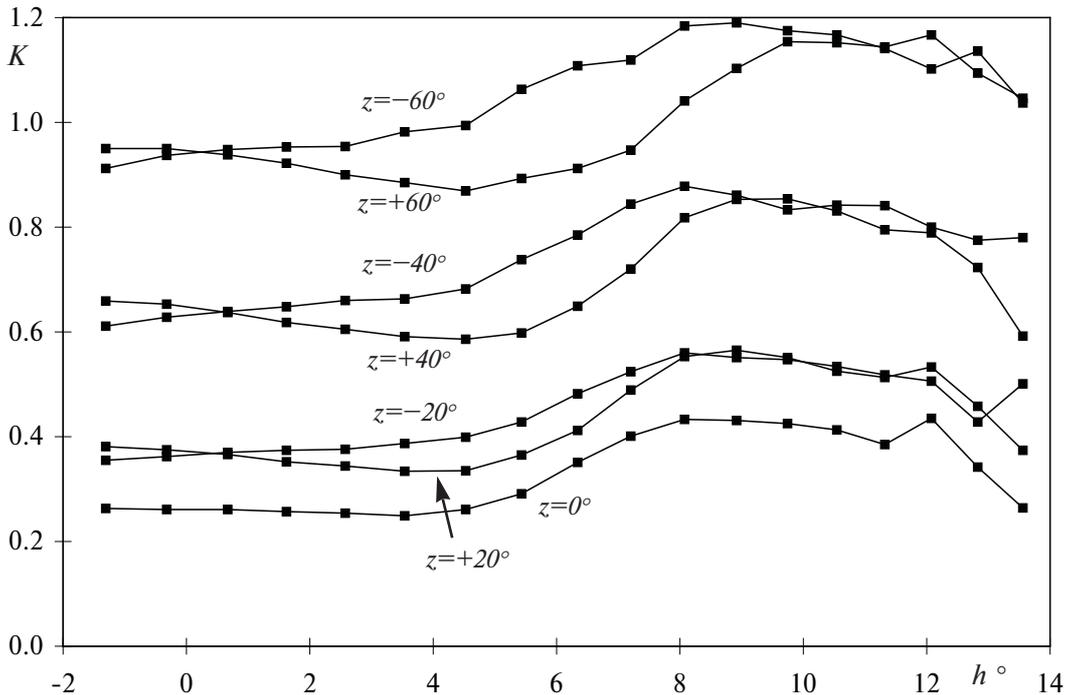

**Figure 5.** Polarization ratio *K* of the twilight sky depending on *h* for different solar vertical points (the evening of 31.VII.1997, 356 nm).

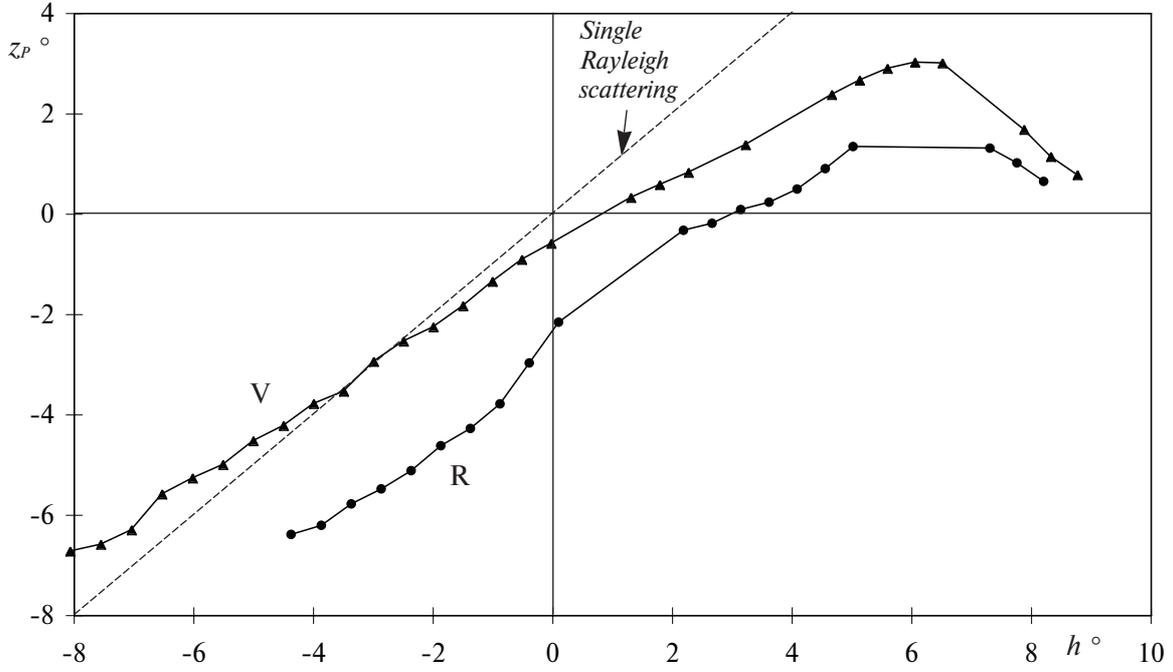

**Figure 6**. The dependence of maximum polarization point zenith distance $z_P$ on $h$ for the simultaneous V and R observations in the evening, 6.VIII.2000.

The effect of polarization changes near the zenith point will be the basic for the method of single and multiple scattering separation that will be suggested in the next part of this work.

## 4. SEPARATION METHOD

Let us consider the case $z=h=0$ and in the case when the influence of single aerosol scattering is small relatively single Rayleigh scattering. Let $A$ be fraction of single scattered light in the total sky brightness for the perpendicular polarization direction, and $Q$ and $q$ be the polarization ratios for single-scattered and multiple-scattered light, respectively. It can be shown that the total polarization ratio is

$$K = AQ + (1-A)q \qquad (7).$$

We make the mixed derivative of $K$ with respect to $z$ and $h$ and express it in terms of the derivatives of $A$, $Q$ and $q$:

$$\frac{d^2K}{dzdh} = A\frac{d^2(Q-q)}{dzdh} + \frac{dA}{dz}\frac{d(Q-q)}{dh} +$$
$$+ \frac{dA}{dh}\frac{d(Q-q)}{dz} + (Q-q)\frac{d^2A}{dzdh} + \frac{d^2q}{dzdh} \qquad (8).$$

Since $\gamma=90°-z$ at $h=0$, the quantity $dQ/dz$ ($z=h=0$) vanishes. It follows from (5), with allowing for $\theta=1$ that $dq/dz=0$ for any $h$. On the other hand, the ratio $I_\perp/f$ at $h=0$ has a minimum at the zenith; therefore, taking in account the constant value of $J_\perp/f$ [as follows from formula (3)] we have $d(j_\perp/f)/dz=0$ and, hence, $dA/dz=0$ for $z=h=0$. As a result, we obtain

$$\frac{d^2K}{dzdh} = A\frac{d^2Q}{dzdh} + (Q-q)\frac{d^2A}{dzdh} \qquad (9).$$

The mixed derivative in first term can be calculated by the direct substitution of (3) (on account of $\gamma=90°-z+h$). It turns out to be equal to $-2/(1+\alpha)$. The second term in the right side of equation (9) also contained a mixed derivative that can be transformed as follows:

$$\frac{d^2A}{dzdh} = \frac{d^2}{dzdh}\left(\frac{J_\perp}{J_\perp + j_\perp}\right) =$$
$$= \frac{d}{dz}\left(\frac{(j_\perp/f)(dJ_\perp/fdh) - (J_\perp/f)(dj_\perp/fdh)}{(J_\perp/f + j_\perp/f)^2}\right) \qquad (10).$$

In the last equality we divided the numerator and denominator by $f^2(z)$. Since the derivatives of the functions $J_\perp/f$ and $j_\perp/f$ with respect to $z$ vanish, the equation takes the form

$$\frac{d^2A}{dzdh} = \frac{j_\perp f}{(J_\perp + j_\perp)^2}\frac{d^2(J_\perp/f)}{dzdh} \qquad (11).$$

The mixed derivative of $J_\perp/f$ can be calculated by the direct substitution of integral (1) or with help of "twilight layer" model. Taking into account that the atmosphere at altitudes between 10 and 30 km is virtually isothermal, with a temperature $T=220K$, and substituting the corresponding Boltzmann distribution function into $n(H)$ we obtain, after simple calculations:

$$\frac{d^2A}{dzdh} = \left(\frac{H_{L0}}{H_{atm}} - 1\right)A(1-A) \qquad (12),$$

where $H_{atm}=6.43$ km is the height of the uniform atmosphere corresponding to the above temperature. Because of assumed lack of aerosol scattering, we can also calculate the quantity $H_{L0}$ from the gas atmosphere model. Next, we use formula (7) to express $(Q-q)$ in terms of

$K$ and $A$ and, substituting (12) into (9), obtain the expression for $A$ at $z=h=0$:

$$A = -\frac{d^2K}{dzdh}\left(\frac{2}{1+\alpha} + \left(K - \frac{\alpha}{1+\alpha}\right)\left(\frac{H_{L0}}{H_{atm}} - 1\right)\right)^{-1} \quad (13).$$

Here, the quantity of $K$ and its mixed derivative at the point $z=h=0$ are determined from the observations. The resulting value is positive because the mixed derivative of $K$ is less than zero. Basing on this value, we can calculate the ratio of single and multiple scattered light at the moment $h=0$ for any polarization direction and amounts of $z$ where measurements were made, by using formulae (2) and (3). And certainly, given the altitude atmospheric model, we can expand the calculations of $A$ for any amount of $h$ in observed range by single scattering calculation, but in this case we shall calculate only Rayleigh fraction of single scattering without taking into account the upper atmospheric aerosol the influence of which at larger $z$ can be sufficient (but, as we shall see below, it did not exceed several dozens of percent at 356 nm for 1997 observation dates).

## 5. RESULTS

The method discussed above gives the exact values of single- to multiple-scattered light ratio only if the influence of single aerosol scattering at the moment $h=0$ (what is corresponding to aerosol scattering at altitudes about 15-25 km) is insufficient. One can expect that it will be true at the clear sky at short wavelengths but this rule should be broken at larger ones. As we saw above, the aerosol scattering was insufficient for the most part of 1997 observations at 356 nm. To check this assumption and formula (13) correctness for all other observations at different wavelengths we build the diagram of two measured quantities which are included to this formula, $K$ and its mixed derivative with respect to $z$ and $h$ $(z=h=0)$, for all observations when these quantities were measured with good accuracy. This diagram is shown in the Figure 7, the cross symbol in this diagram corresponds to the case of clear Rayleigh single scattering, when

$$K = Q = \frac{\alpha}{1+\alpha} \cong 0.03; \quad \frac{d^2K}{dzdh} = \frac{-2}{1+\alpha} \cong -1.94 \quad (14).$$

We can see that most part of points corresponding to the 1997 observations at 356 nm and also most part of points corresponding to the 2000 observations in U, B and V bands are situated near the straight line passing also near the point of clear Rayleigh single scattering. But most part of points corresponding to the observations in R band, where aerosol influence is expected to be most sufficient, deflect from this line in the same direction and only one R band point (evening of August, 4th, 2000) is situated near the line. Here we should notice that in that evening the weather conditions were the best and the sky polarization was the strongest for all V and R bands observation dates.

Taking these facts into account we may assume that single aerosol scattering at the sunrise and sunset was insufficient for the most part of 356 nm, U, B and V observations, with the exception of morning of July, 30th, 1997 (356 nm) and morning of the same date in 2000 (B band). Small influence of aerosol was also possible in the mornings of July, 28th, 2000 (B band) and August, 5th-7th, 2000 (V band). All these points deflect from the line at the same direction as the points of R-band observations, where the aerosol influence is sufficient. It is interesting to notice that single aerosol scattering at 356 nm, U, B and V bands was seen only in morning twilight observations.

Confirming the fact above, we should add that linear correlation between the values of $K$ and its mixed derivative in the case of lack of single aerosol scattering quite naturally results from (7) and (13) with account of constant value of $\alpha$, small value of second item (relatively to the first one) in the brackets of formula (13) and weak dependence of quantity $(Q-q)$ on the wavelength what is expected in the case of Rayleigh scattering.

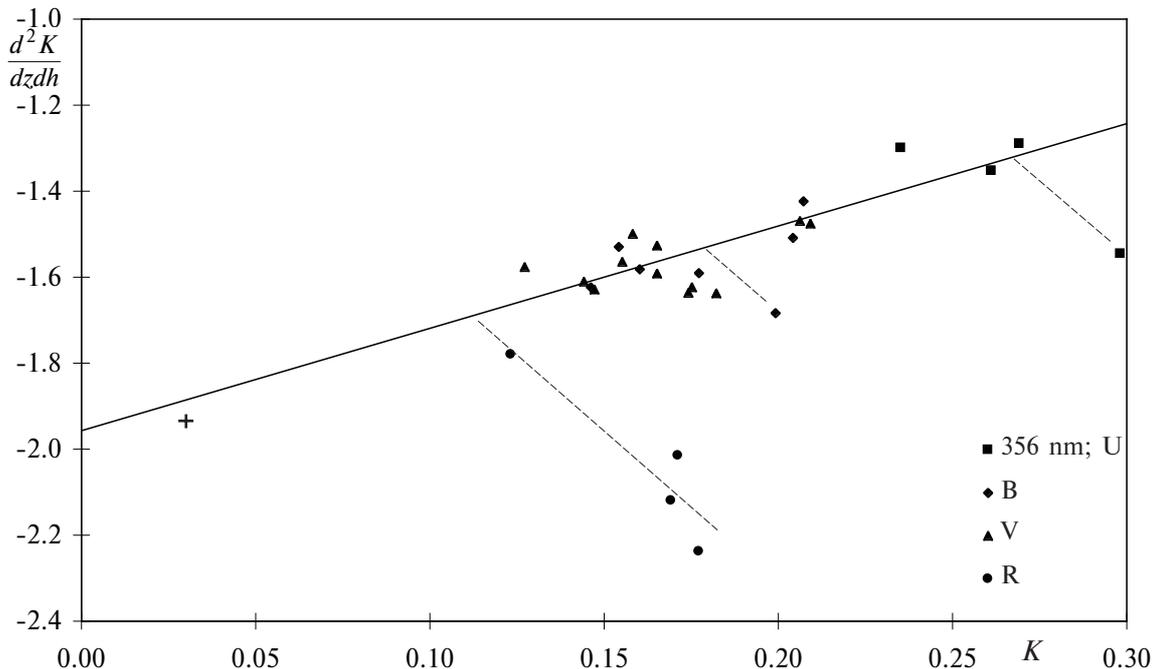

**Figure 7.** The correlation of $K$ and its mixed derivative for 1997 and 2000 observations.

We should not conclude that atmosphere aerosol does not display itself at clear twilight sky in the blue-green part of spectrum. We can speak only about small value of its scattering coefficient relatively Rayleigh one during the most part of observation dates and only at scattering angles about 90° for these wavelengths and effective scattering altitudes for the sunrise and sunset time. For our color bands these altitudes are about 15-25 km. Naturally, the influence of aerosol scattering sufficiently increases in R band where the ratio of aerosol and Rayleigh scattering coefficients increases and the effective scattering takes place at the lower altitudes with large density of aerosol particles.

However, our conclusion is enough for possibility to calculate the contribution of single scattering at $z=h=0$ by formula (13) in the most part of our observations at 356 nm and in U, B and V bands. The quantity $H_{L0}$ was calculated based on the gas atmosphere model with temperature distribution typical for the season and latitude of observations (it is shown in Figure 8). Its value turned out to be 24.6 km for 356 nm, 24.0, 18.9 and 13.4 km for U, B and V bands, respectively, in the last case the ozone Chappuis absorption was taken into account.

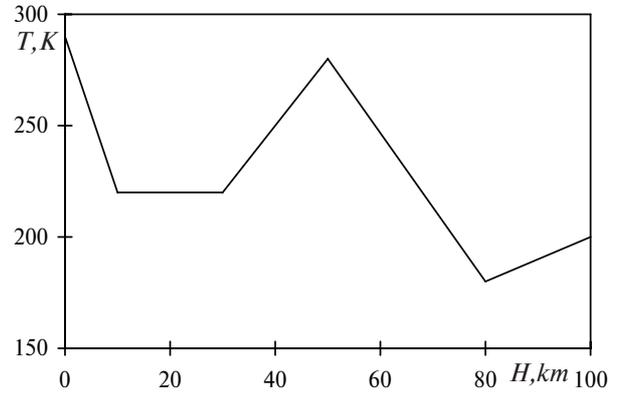

**Figure 8.** Model temperature distribution in the atmosphere.

Table 1 contains the results for 1997 and 2000 observation dates: the value of $K(z=h=0)$, its mixed derivative with respect to $z$ and $h$, the contribution of single scattering in the twilight sky background at the zenith at $h=0$ for polarization plane perpendicular (A) and parallel (a) to scattering plane and for total sky background $(A_T)$. The star symbol (*) near these values means their possible inaccuracy due to the presence of single aerosol scattering. Last column of the table contains relative error of mixed derivative of $K$ that is almost coincides with relative errors of $A$, $a$, and $A_T$.

| Color band or $\lambda$ | Observation date | $K$ | $\dfrac{d^2 K}{dzdh}$ | $A$ | $a$ | $A_T$ | $E$ |
|---|---|---|---|---|---|---|---|
| | **1997** | | | | | | |
| 356 nm | 30.VII, the morning | 0.298 | −1.544 | 0.572* | 0.056* | 0.441* | 0.005 |
| | 31.VII, the evening | 0.261 | −1.351 | 0.521 | 0.058 | 0.413 | 0.005 |
| | 1.VIII, the morning | 0.269 | −1.288 | 0.492 | 0.053 | 0.388 | 0.010 |
| | **2000** | | | | | | |
| U | 17.VII, the evening | 0.235 | −1.298 | 0.518 | 0.064 | 0.432 | 0.013 |
| B | 25.VII, the evening | 0.207 | −1.422 | 0.622 | 0.088 | 0.530 | 0.023 |
| | 26.VII, the morning | 0.204 | −1.507 | 0.661 | 0.094 | 0.565 | 0.037 |
| | 28.VII, the morning | 0.177 | −1.589 | 0.713* | 0.118* | 0.624* | 0.011 |
| | 28.VII, the evening | 0.154 | −1.528 | 0.700 | 0.132 | 0.624 | 0.007 |
| | 29.VII, the morning | 0.146 | −1.622 | 0.748 | 0.149 | 0.672 | 0.011 |
| | 29.VII, the evening | 0.160 | −1.580 | 0.720 | 0.131 | 0.639 | 0.013 |
| | 30.VII, the morning | 0.199 | −1.682 | 0.740* | 0.108* | 0.635* | 0.031 |
| V | 30.VII, the evening | 0.206 | −1.468 | 0.688 | 0.097 | 0.587 | 0.022 |
| | 31.VII, the evening | 0.209 | −1.474 | 0.690 | 0.096 | 0.587 | 0.016 |
| | 3.VIII, the evening | 0.155 | −1.563 | 0.752 | 0.141 | 0.670 | 0.019 |
| | 4.VIII, the morning | 0.147 | −1.627 | 0.786 | 0.156 | 0.705 | 0.020 |
| | 4.VIII, the evening | 0.127 | −1.575 | 0.769 | 0.177 | 0.702 | 0.013 |
| | 5.VIII, the morning | 0.174 | −1.635 | 0.779* | 0.130* | 0.682* | 0.032 |
| | 5.VIII, the evening | 0.144 | −1.609 | 0.779 | 0.157 | 0.700 | 0.025 |
| | 6.VIII, the morning | 0.182 | −1.636 | 0.776* | 0.124* | 0.676* | 0.031 |
| | 6.VIII, the evening | 0.165 | −1.525 | 0.730 | 0.129 | 0.645 | 0.019 |
| | 7.VIII, the morning | 0.175 | −1.622 | 0.772* | 0.128* | 0.676* | 0.021 |
| | 7.VIII, the evening | 0.158 | −1.498 | 0.720 | 0.133 | 0.639 | 0.023 |
| | 8.VIII, the morning | 0.165 | −1.590 | 0.761 | 0.135 | 0.673 | 0.025 |
| R | 4.VIII, the evening | 0.123 | −1.778 | 0.893* | 0.211* | 0.818* | 0.049 |
| | 6.VIII, the evening | 0.169 | −2.118 | — | — | — | 0.033 |
| | 7.VIII, the evening | 0.177 | −2.236 | — | — | — | 0.033 |
| | 8.VIII, the evening | 0.171 | −2.013 | — | — | — | 0.012 |

**Table 1.** The single scattering contribution at $z=h=0$ for 1997 and 2000 observation dates.

As we can see from the table, at $z=h=0$ single scattering contains less than the half of total sky background brightness in a violet part of spectrum but with increase of wavelength its contribution becomes more sufficient and can reach 70% in a green-yellow part of spectrum. Last result is quite near to theoretical estimations of this value [2], but does not reach them. We can conclude that multiple scattered component of the twilight sky is "more blue" than single scattered one that is quite natural if we take into account the increase of scattering coefficient at the short wavelengths.

The contribution of single scattering was changing from day to day, notice its sufficient decrease in the evenings of July, 30th and 31th, 2000. One of the possible explanations of this phenomenon is so-called "barrier effect" which was considered in [2] and displayed in [7] as sufficient decrease of sky brightness in long-wave part of spectrum which was not related with atmosphere conditions changes directly at the observation point. This lead to the change of sky color clearly visible for two lowest curves in Figure 3. The scheme of this effect is shown in Figure 9. If the barriers for light propagation (clouds) appear at the direction of the setting sun, it can sufficiently decrease the brightness of single-scattered light, especially at long-wave part of spectrum, where the "twilight ray" is propagating at low altitudes. The influence of the barrier on the multiple scattered component is not so sufficient, and its contribution increases. This picture is also confirmed by the clouds appeared at the glow region after the sunset in these two evenings.

Another one possible reason for changes of single- and multiple-scattered light ratio is the changes of lower tropospheric aerosol that is not emitted by the sun and does not form the single scattered light but have influence on multiple scattering and absorbs all components of twilight sky background.

The picture in the R color band is more complicated, as it was mentioned above. Strong influence of atmospheric aerosol for most part of observations moves the points down and right in the Figure 7. The module of mixed derivative of $K$ sufficiently increases, although the value of $K$ itself only slightly differs from that one in the V color band. Formal substitution of the obtained parameters into formula (13) in some cases leads to the value of $A$ exceeding the unity, that is obviously unreal (these cases are corresponded by the dashes in Table 1).

The reasons of $K$ mixed derivative module increase can be related with the polarization properties of aerosol scattering and with fast decrease of aerosol scattering coefficient with the altitude, that agrees with its weak influence at larger altitudes that are effective for scattering in the V band. This effect would mean decrease of $H_{atm}$, increase of the second item module in formulae (9) and (13) and, thus, increase of $K$ mixed derivative module.

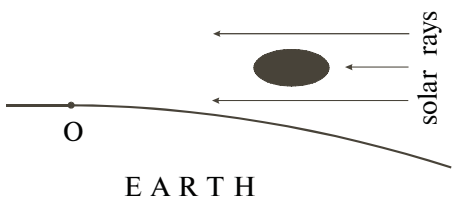

**Figure 9.** The "barrier effect" decreasing the brightness of single-scattered light.

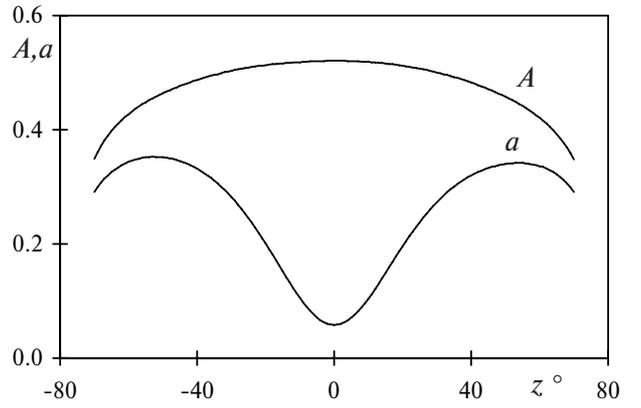

**Figure 10.** Single-scattered light contribution at $h=0$ for two polarization directions (the evening, 31.VII.1997, 356 nm).

All these reasons do not allow us to use the formula (13) to estimate the contribution of single scattering for the R color band in general. However, approximate estimations can be made for the evening of August, 4th, 2000, when the influence of aerosol was quite weak. Assuming $H_{L0}=9.8\ km$ (from the gas atmosphere model), we obtain the value of $A$ given in Table 1. We can see that the contribution of single scattering continue to rise with the increase of wavelength from V to R band.

Having obtained the contribution of single scattering at the point $z=h=0$ we can estimate it for the other cases in the observations range. Figure 10 shows the dependence of this value on $z$ at $h=0$ for two polarization directions for the evening of July, 31th, 1997 (356 nm). Zenith minimum for parallel polarization direction appears due to the strong polarization of Rayleigh scattering at $\gamma=90°$.

As it was mentioned above, in the case of $h>0$ we can calculate only Rayleigh fraction of single scattering, assuming the altitude temperature distribution shown in Figure 8 (the value of temperature of mesospheric minimum region, 80 km, is taken from [9]). The result of calculations for the same evening are shown in the Figure 11. An interesting fact is worthy of note: at the bright period of twilight, the contribution of single scattering slightly increases with the sun's depth under the horizon for some zenith distances. But at $h>5°$ this value rapidly decreases, especially in the opposite to the glow sky region, where the effective scattering altitude rises faster than in the glow region of the twilight sky. These two twilight stages were called in [2] "half-twilight" and "total twilight".

Let us return to the formula (4) for multiple scattering. As it was mentioned above, this is only an approximate formula, and its integration gives rise to perceptible errors. Nevertheless, we can rewrite (4) as

$$j(z,h) = r(z,h)j(-z,h) \qquad (15),$$

where $r(z,h)$ is a slowly varying function of $h$ which depends parametrically on $z$. The similar function for single scattering

$$R(z,h) = \frac{J(z,h)}{J(-z,h)} \qquad (16)$$

exhibits quite different properties: owing to the variance of effective scattering altitudes, it rises sharply with increasing $h$. The total brightness ratio $R_0(z,h)$ is connected with these functions by a relation similar to (7) for polarization ratio:

$$R_0(z,h) = \frac{I(z,h)}{I(-z,h)} = A(-z,h)R(z,h) + \\ + (1 - A(-z,h)) r(z,h) \quad (17).$$

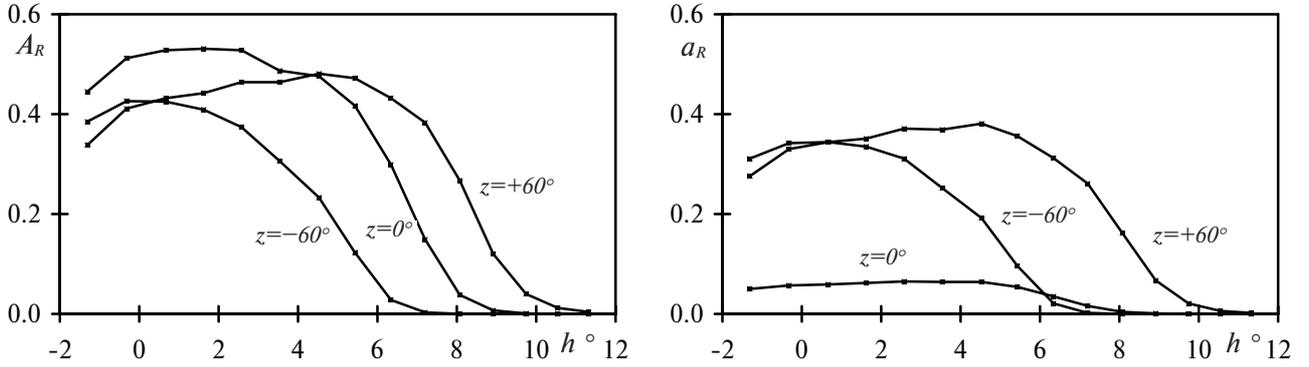

**Figure 11.** The contribution of Rayleigh single scattering depending on $h$ for two polarization directions and different values of $z$ (the evening, 31.VII.1997, 356 nm).

The upper curve in Figure 12 shows this dependence for the evening of July, 31[th], 1997 (356 nm) and polarization direction perpendicular to the scattering plane, $z=60°$. It is apparent that for $h=6°$ the growth of $R_0$ changes to a decrease and at $h=10°$, the function $R_0(h)$ reaches a plateau, where its value is slightly higher than the value at $h=0$. Taking into account the properties of functions $R(h)$ and $r(h)$, this points unambiguously to the fact that multiple scattering is completely dominant for $z=60°$ (in the glow region) at $h>10°$. Correspondingly, this situation occurs earlier at the zenith region (as it visible in the Figure 7, for example) and much earlier — on the opposite to the glow side of the sky. The role of single scattering becomes disappearingly small at effective scattering altitudes of 100-120 km.

To estimate the possible aerosol scattering coefficient at high altitudes (corresponding to $h>0°$), we plot a similar dependence after subtracting the Rayleigh component calculated above from the total sky background (lower curve in the same figure). At this case we shall also find a peak at $h=6°$, though with smaller height. This gives evidence of the presence of "nonsubtracted" single-scattered light (obviously scattered by aerosol particles). But comparing the picks magnitudes we can conclude that aerosol single scattered component does not exceed several dozens of percents of Rayleigh one (at least for this observation date). Thus, graphs presented in Figure 11, enable us to estimate the role of single scattering in general.

To summarize the review of observational data, we show in Figure 13 the polarization ratios $q$ for multiple-scattered light as a function of $z$ for $h=0°$ and $h=12°$ for the same observation date (evening of July, 31[th], 1997, 356 nm). These curves, differing a bit from each other, satisfy equation (5) with $\theta=1$ with good accuracy, and agree reasonably with the model curve constructed for double scattering under the assumption that the secondary light source is a thin glowing semicircle with a gradual fall of intensity from the center (above the sun) toward the edges. The multiple scattering becomes unpolarized at zenith distances about 50°, and these "neutral points" of multiple scattering, as it is visible in the figure, are practically immovable during the twilight period! So, the motion of "neutral points" for total background investigated in a large number of papers [2] can be considered simply as the reflection of single- to multiple-scattering ratio change.

At the zenith distances more than 50°, the "inverse polarization" effect takes place ($j_\parallel > j_\perp$), which is caused by the scattering of light coming from the side (relative to the solar vertical) regions of the sky sphere. This effect, such as the effect of "neutral points" motion completely described by the influence of multiple scattering and do not related with atmospheric aerosol as it was assumed in [2].

## 6. DISCUSSION

The main result of this work is the estimate of the fraction of multiple-scattered light in the total background of the twilight sky for various wavelengths, depths of the sun under horizon and zenith distances of the sky point in the solar vertical. For most part of 1997 observations at 356 nm and 2000 observations in U, B and V color bands the influence of aerosol scattering had turned out to be insufficient (except a little number of morning observations only) that gave us the possibility of exact estimation of the single scattering contribution. In the R color band the aerosol scattering becomes more sufficient, but the polarization is even stronger than in V band, that points to the strong polarization of aerosol scattering. The R band data also shows that aerosol scattering coefficient is rapidly decreasing with the altitude.

The contribution of single scattered light in the total twilight sky background at the zenith at sunrise or sun-set moment is equal to only about 40% for the violet rays, but increases to 60% in blue part of spectrum and sometimes reaches 70% for visible one. Strong aerosol influence does not allow to estimate this value for red part of spectrum, but it is expected to be not less than 80%. Such wavelength dependence shows that multiple scattered light has the surplus in short-wave part of spectrum relatively single-scattered one. Relative weak wavelength dependence between B and V bands is possibly owing to sufficient Chappuis ozone absorption of single scattered light in V band.

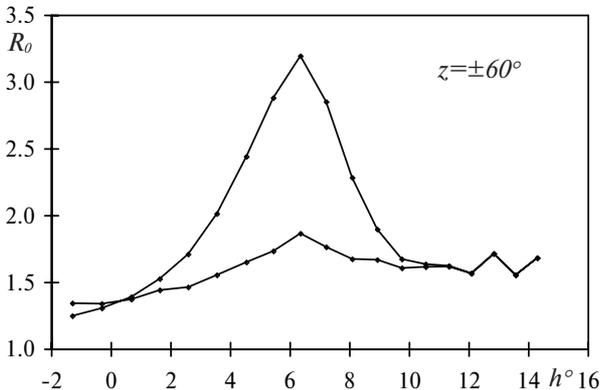

**Figure 12.** Brightness ratio in symmetrical points of solar vertical depending on $h$ (the evening, 31.VII.1997, 356 nm).

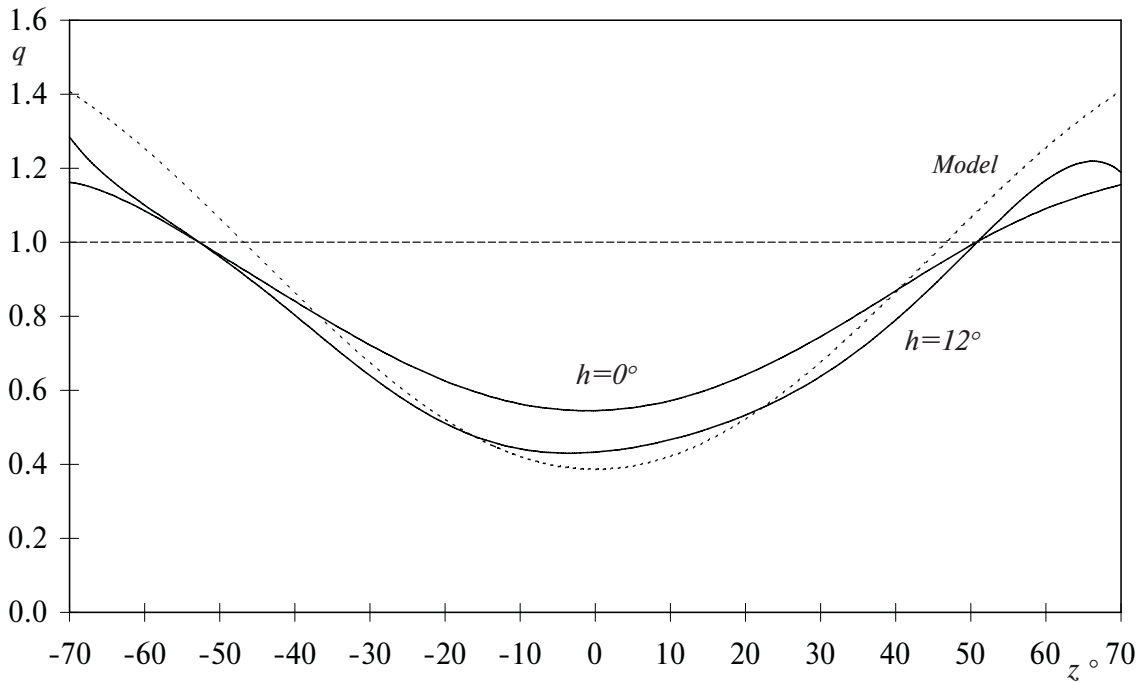

**Figure 13.** Polarization ratio of multiple-scattered light depending on $z$ for the evening, 31.VII.1997, 356 nm.

This ratio remains almost constant until the depth of the sun under horizon about 4-5°, then the role of single scattering rapidly decreases, which is showed by the return of maximum polarization point to the zenith (Figure 7) and weakening of glow surplus of the sky brightness (Figure 12). At $h$ about 10° single-scattered light completely disappears on the multiple-scattered background.

The fraction of multiple scattering, especially at large values of $h$, turned out to be much higher than most of existing estimates [2]. The essence of the problem is the underestimation of the scattering of third and higher orders in theoretical works. The mere fact of the predominance of multiple scattering even during bright twilight period suggests the essential role of high-order scattering. This fact can also be the reason of incorrect dependence of multiple-scattered light on the wavelength obtained in [2,10].

G.V. Rozenberg's book [2] also contains a reference to another work devoted to multiple scattering and published as early as the 1930[th] [11]. The author, Hulburt, attempted to estimate the brightness of multiple scattering at the zenith by representing it as the product of air density at altitudes from 15 to 40 km and the sky brightness measured in the glow region. Despite some inaccuracies committed in this work and noticed later [2], the idea by itself was correct, because method was designed to take scattering of all orders into account. It is quite logical that Hulburt came to the right conclusion that at $h=10°$, single scattering is lost on the background of multiple scattering. It is the conclusion that has become the object of strong criticism in [2] and many other works based on theoretical estimates of the contribution of multiple scattering.

We should also mention the method of separation of single- and multiple-scattering light components that is being developed at the Astronomical Observatory at Odessa University (its basic principles are described in [5]). The essence of this method is an extrapolation of the empirical property of multiple scattering (the constancy of the logarithmic derivative of its brightness with respect to $h$), which was revealed in the region of large $h$ to the region of small $h$. The method is attractive in its simplicity, but, unfortunately, the multiple scattered component does not follow this property very well, showing a significant deviation at $h=8-9°$, just where the single scattering in the mesosphere is still noticeable. In our opinion, this deviation can be accounted for by rapid change of the contribution of double scattering against the background of higher-order scattering. As a result, the scattering coefficient turns out to be overstated, particularly at high scattering angles, which makes the study of mesospheric aerosol difficult.

The main features of multiple scattered light (relative weak polarization with $q$ about 0.5 and surplus at blue part of spectra) easily describe the correlation of color and polarization changes of the twilight sky while the sun is depressing under horizon. The explanation of these effects with the account of only single scattering meets the difficulties.

A number of color measurements of the sky ([2,7], see Figure 3) and polarimetric measurements made here reveal the simultaneous change of the sky color and polarization. The sky reddening is accompanied by increasing of polarization ($K$ decrease) and vice versa. This evidently shows that these changes can have the common reason — change of single- to multiple-scattered light ratio.

At the light stage of twilight which starts before the sunset and continues until the depth of the sun under horizon about 4-5°, observations [7] showed the reddening of the sky, and our 1997 and 2000 observations revealed the slow increase of polarization degree. The role of "red" and more polarized single scattered light at this stage even slightly increase with the depression of sun. Besides that, the color of the sky at this stage can be affected by ozone Chappuis absorption and "barrier effect", as it was noticed in [7], but these effects were overestimated due to multiple scattering neglecting in that work.

When the sun's depth under horizon exceeds 4-5°, the picture sufficiently changes — the sky gets bluer and its polarization falls. No doubt that these effects are related with the rapid decrease of single scattered light contribution. But the polarization of multiple scattered component itself slowly increase with the depression of Sun, as we can see in Figure 13, due to the fact that the secondary source of light (the glow segment) becomes more narrow. Finally, at sun's depth under horizon about 8-10°, when the single scattering completely disappears, the polarization decrease (the increase of $K$) stops and polarization increase starts.

Rapid overestimation of single scattered light contribution made G.V. Rozenberg to relate this polarization minimum at $h=10°$ with the existence of steady aerosol layer at the altitudes of 80-100 km (in the mesosphere) [2], however, this assumption accompanied by the blue surplus of the sky spectrum may lead to the truly fantastic properties of such aerosol particles: their scattering coefficient decreases with the wavelength faster than the Rayleigh scattering law ($\lambda^{-4}$) [12]! It seems quite logical that the same journal release contained the paper [13] where the polarization properties in blue part of spectra for this stage of twilight were related with multiple scattering, although for larger wavelength the influence of mesospheric aerosol was permitted.

All further changes of the twilight sky and polarization at the dark twilight stage ($h>10°$), in particular, its next reddening noticed in [2] can be related only with the properties of multiple scattered component and appearing nighttime sky background, that leads to polarization decrease at the threshold of twilight turn to night.

## 7. CONCLUSION

The primary goal of these work was to develop the method of separation of single- and multiple-scattered light in the twilight sky background and its use for the results of polarimetric twilight observations conducted in 1997 and 2000 for different color bands.

This method separates two components with a good accuracy if the influence of atmospheric aerosol single scattering is insufficient. This situation took place for most part of observations at the 356 nm wavelength and in U, B, and V color bands. The exception was only some number of morning observations. The data obtain in the R color band shows strong influence of atmospheric aerosol and we have found only approximate value (more possibly, the lower limit) of single scattering contribution.

The results being obtained shows the increase of single scattering contribution in long-wave part of spectrum. This fact along with the polarization properties of multiple scattering explains the color and polarization changes of the twilight sky with the sun's depression under horizon. The values of single scattering contribution are near to theoretical ones only for light stage of twilight, it becoming less than theoretical ones at the deeper stages. The reasons for these disagreements are discussed in this paper.

The increase of single scattered light contribution with the wavelength makes the green-yellow part of spectrum (the V band) to be the most effective for twilight probing the atmosphere and its ozone layer by Chappuis absorption. Further increase of wavelength will meet the difficulties related with tropospheric aerosol light scattering.

ACKNOWLEDGMENTS

The work was made with financial assistance of Russian Foundation for Basic Research, grants 00-02-16396, 01-02-06247, 01-05-99403. Authors would like to thank Yu. I. Zaginailo, I.V. Zaginailo, V.D. Motrich, S.B. Yudin, V.I. Shenavrin, P.A. Prudkovskii and P.D. Zhuravlev for their help in observations and laboratory investigations of devices.